\documentclass[10pt]{iopart}		

\usepackage{iopams}  

\usepackage{graphicx}

\usepackage[T1]{fontenc} 
\usepackage{multirow}
\usepackage{color}
\usepackage[normalem]{ulem}

\begin{document}

\title{Phase engineering in tantalum sulfide monolayers on Au(111)}

\author{Daniela Dombrowski$^{1,2}$, Abdus Samad$^3$, Kai Mehlich$^{1,2}$, Thais Chagas$^1$, Udo Schwingenschl\"ogl$^3$, Carsten Busse$^{1,2}$}

\address{$^1$ Department Physik, Universit\"at Siegen, Germany}
\address{$^2$ Institut f\"ur Materialphysik, Westf\"alische Wilhelms-Universit\"at M\"unster, Germany}
\address{$^3$ Physical Science and Engineering Division, King Abdullah University of Science and Technology (KAUST), Thuwal 23955-6900, Saudi Arabia}

\ead{carsten.busse@uni-siegen.de}

\vspace{10pt}
\begin{indented}
\item[]\today
\end{indented}

\begin{abstract}
We prepare monolayers of tantalum sulfide on Au(111) by evaporation of Ta in a reactive background of H$_2$S. Under sulfur-rich conditions, monolayers of 2H-TaS$_2$ develop, whereas under sulfur-poor conditions TaS forms, a structure that can be derived from 2H-TaS$_2$ by removal of the bottom S layer. We analyse the alignment of the layers with respect to the substrate and the relation with the domains in the Au(111) herringbone reconstruction using scanning tunneling microscopy (STM). With the help of density functional theory (DFT) calculations we can determine the registry of the two phases with the substrate. We develop a growth process that allows preparation of uniquely oriented 2H-TaS$_2$ on Au(111). 2H-TaS$_2$ and TaS have a remarkably similar in-plane lattice structure and we observe the formation of lateral 2H-TaS$_2$-TaS heterostructures with atomically well-defined and defect-free boundaries. We observe mirror twin boundaries within 2H-TaS$_2$ along the S- and Ta-edge.
\end{abstract}

%
%
%
%
\ioptwocol


\section{Introduction}

The properties of ultrathin materials are determined by composition, phase, and film thickness. Phase engineering is the rational use of the same ingredients to obtain different materials with varying properties, and potentially a powerful tool to tune unique functionalities and innovative properties of distinct systems according to the intended applications and, consequently, to enable their incorporation in future devices \cite{Chen2020}.

Ultrathin films with the same ingredients can have distinct properties depending on their stoichiometry and atomic coordination \cite{Wang2018}. An especially rich field for phase engineering are materials (formally) derived from transition metal dichalcogenides (TMDCs) of the form MX$_2$ (M = metal, X = chalcogen). Different phases can be favored depending on the growth conditions and modified through post-growth annealing processes \cite{Bonilla2020}. Recently, there were several studies on chalcogen-poor variants of TMDCs, both in the form of self-intercalated layers \cite{Zhao2020} or monolayers with an ordered arrangement of chalcogen vacancies \cite{Arnold2018,Cheng2018}.
 
The reduced symmetry is decisive for key properties of ultrathin TMDCs. Specifically, the loss of inversion symmetry when going from the bilayer to the monolayer in the 2H-phase \cite{Chhowalla2013} can directly lead to a spin-polarized electronic band structure. For example, in monolayer 2H-TaS$_2$ the band crossing the Fermi level is spin-split in a large fraction of the Brillouin zone \cite{Sanders2016}. For fundamental reasons, the spin-splitting has to be reversed for points in the Brillouin zone linked by inversion symmetry, for example, the K and K' point in a hexagonal crystal lattice, thereby enabling valleytronics \cite{Xiao2012}.

\begin{figure}
	\centering
		\includegraphics[width=\columnwidth]{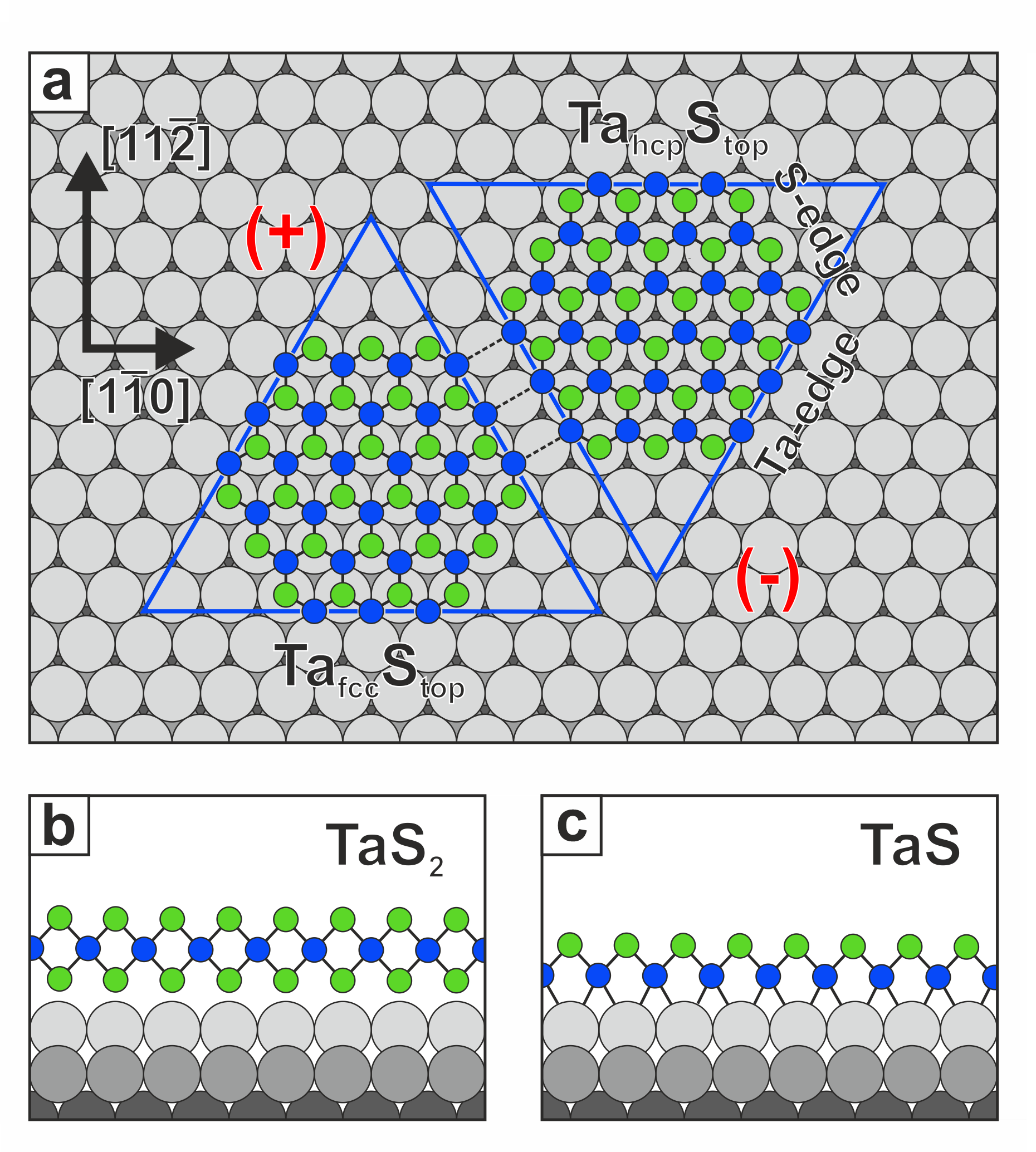}
	\caption{{\bf Registry of TaS$_2$ and TaS on Au(111).} Blue circles: Ta; green circles: S; grey circles: Au (shaded differently in different layers). \textbf{(a)} Top view for both TaS$_2$ (green circle denotes two S on top of each other) and TaS (green circle denotes single S). Ta$_{\rm fcc}$S$_{\rm top}$ and Ta$_{\rm hcp}$S$_{\rm top}$ label the two possible registries, see text. (+) and (-) label the two resulting orientations. The two different edge types are labelled Ta-edge and S-edge, see text. Blue triangles indicate the triangular form in case of preference for Ta-edges. Dashed lines symbolize bonding across a mirror twin boundary (MTB). Arrows indicate crystal directions. \textbf{(b)} Side view of a thin slice cut out perpendicular to the $[11\bar{2}]$-direction for TaS$_2$ and \textbf{(c)} for TaS.}
	\label{fig:TaS2Au111stackings}
\end{figure}

Our study focuses on TMDCs composed of tantalum and sulfur. This system has a rich phase diagram in the bulk \cite{Jellinek1962} which opens the door to the tuning of properties without changing the ingredients. A versatile method for the preparation of well-defined ultrathin layers of tantalum sulfide is epitaxial growth on single-crystalline substrates \cite{Sanders2016}. In our previous work \cite{Dombrowski2021} we discovered a sulfur-poor tantalum sulfide phase coexisting with the previously established monolayer 2H-TaS$_2$. The 2H-phase is schematically shown in figures~\ref{fig:TaS2Au111stackings} (a) and (b). It has a honeycomb lattice where one sublattice is occupied by Ta and the other by two S on top of each other. Note that in the top view shown in figure~\ref{fig:TaS2Au111stackings} (a) the lower S is eclipsed by the upper one. The side view in figure~\ref{fig:TaS2Au111stackings} (b) shows the typical chalcogen-metal-chalcogen sandwich structure of TMDCs. The bonding between 2H-TaS$_2$ and the substrate is mainly given by the van der Waals interaction \cite{Silva2021}. The monolayer 2H-TaS$_2$ phase was called $\delta$-phase in Ref.~\cite{Dombrowski2021}, it will be simply referred to as TaS$_2$ in this paper.

Under S-poor conditions a S-deficient phase can grow ($\beta$-phase in Ref.~\cite{Dombrowski2021}). We determined its stoichiometry as TaS$_{2-x}$ with $0 \leq x \leq 1$. It will be called TaS in this paper to avoid cumbersome notation. When we refer to both phases at the same time we will use the neutral term \emph{tantalum sulfide}. A slightly simplified model of TaS (for a full discussion see Ref.~\cite{Dombrowski2021}) is shown in figure~\ref{fig:TaS2Au111stackings} (a) and (c). In the side view in figure~\ref{fig:TaS2Au111stackings} (c) the lower S layer is missing, the undercoordinated Ta atoms interact strongly with the Au substrate. Conveniently, we can use the same schematic top view (figure~\ref{fig:TaS2Au111stackings} (a)) as for TaS$_2$ when we now interpret the green circle as a single S. 

In this paper, we first describe the structure of both tantalum sulfide phases with special emphasis on the epitaxial relation with the substrate. We present a protocol for the preparation of TaS$_2$/Au(111) with a unique orientation. Finally, we analyse boundaries between different phases and different orientations.


\section{Methods}

We used an ultra-high vacuum system (home-built, background pressure $P = 5 \times 10^{-11}$~mbar). Samples were prepared in-situ, temperatures were measured by thermocouples in direct contact with the samples.

Au(111) was prepared by 1.5~keV Ar$^+$ sputtering at room temperature followed by a high temperature sputter cycle at 900~K and annealing at 900~K. Cleanliness was verified by a well-defined low-energy electron diffraction (LEED) pattern and absence of impurities in scanning tunneling microscopy (STM). 

Ta was evaporated from a mini e-beam evaporator (OAR EGCO4), the evaporation rate $R_{\rm Ta}$ was calibrated by evaporation of Ta on Au(111) at room temperature followed by annealing at 500~K for 10~min leading to pseudomorphic Ta islands of monolayer height, the coverage $\Theta_{\rm Ta}$ of which is used to quantify the amount of Ta. H$_2$S was dosed using a stainless steel tube (tube diameter $d \approx 1$~cm, distance from tube opening to sample $s \approx 3$~cm, angle between tube axis and sample normal $\theta = 50^{\circ}$). The pressure $P_{\rm H_2S}$ was measured far away from the tube opening.

Tantalum sulfide was prepared by evaporation of Ta at a rate $R_{\rm Ta}=0.03$~ML/min at room temperature up to a coverage of $\Theta_{\rm Ta}=0.3$~ML in an H$_2$S atmosphere quantified by $P_{\rm H_2S}=10^{-8}$~mbar, followed by an annealing step for $\approx 25$~min at $T = 700 - 850$~K. During annealing and also until the sample temperature drops below 450~K $P_{\rm H_2S}$ is kept constant. Deviations from these standard parameters are given in the figure captions. A detailed study on the dependence of the distribution and morphology of the tantalum sulfides is given in our previous publication \cite{Dombrowski2021}.

STM measurements were conducted at room temperature. Tunneling parameters (bias voltage $U$, tunneling current $I$) are given for each image. 

To perform spin-degenerate density functional theory (DFT) calculations we use the Vienna \emph{ab initio} simulation package (projector augmented wave method) \cite{PhysRevB.59.1758}. The generalized gradient approximation is applied to the exchange-correlation potential \cite{perdew1996generalized}. The Grimme semi-empirical method is used for the van der Waals correction \cite{grimme2006semiempirical}. We use the valence electron configurations Ta 5$p^6$, 6$s^2$, 6$d^3$, S 3$s^2$, 3$p^4$, and Au 5$d^{10}$, 6$s^1$. An energy cut-off of 360 eV is employed for the plane-wave basis. For the $7{\times}7{\times}1$ supercell of 2H-TaS$_2$ or TaS in contact with an $8{\times}8{\times}1$ Au(111) supercell (4 Au layers), a $2{\times}2{\times}1$ $k$-mesh is used. Slab models are obtained by adding a vacuum layer of 15~{\AA} thickness in the out-of-plane direction. The two Au layers farthest away from the interface are fixed at the atomic positions of bulk Au. The total energy convergence criterion of the self-consistency calculations is set to $10^{-6}$ eV. The atomic positions and lattice constants are optimized until the Hellmann--Feynman forces remain below $10^{-2}$ eV/{\AA} and the remaining pressure is reduced to less than 1 kbar. STM images are simulated by the Tersoff--Hamann method \cite{tersoff1985theory}. Partial charges are calculated using Bader charge analysis \cite{Yu2011}.


\section{Results and discussion}

\subsection{Structure, orientation, stacking, and registry}

\begin{figure}
	\centering
		\includegraphics[width=\columnwidth]{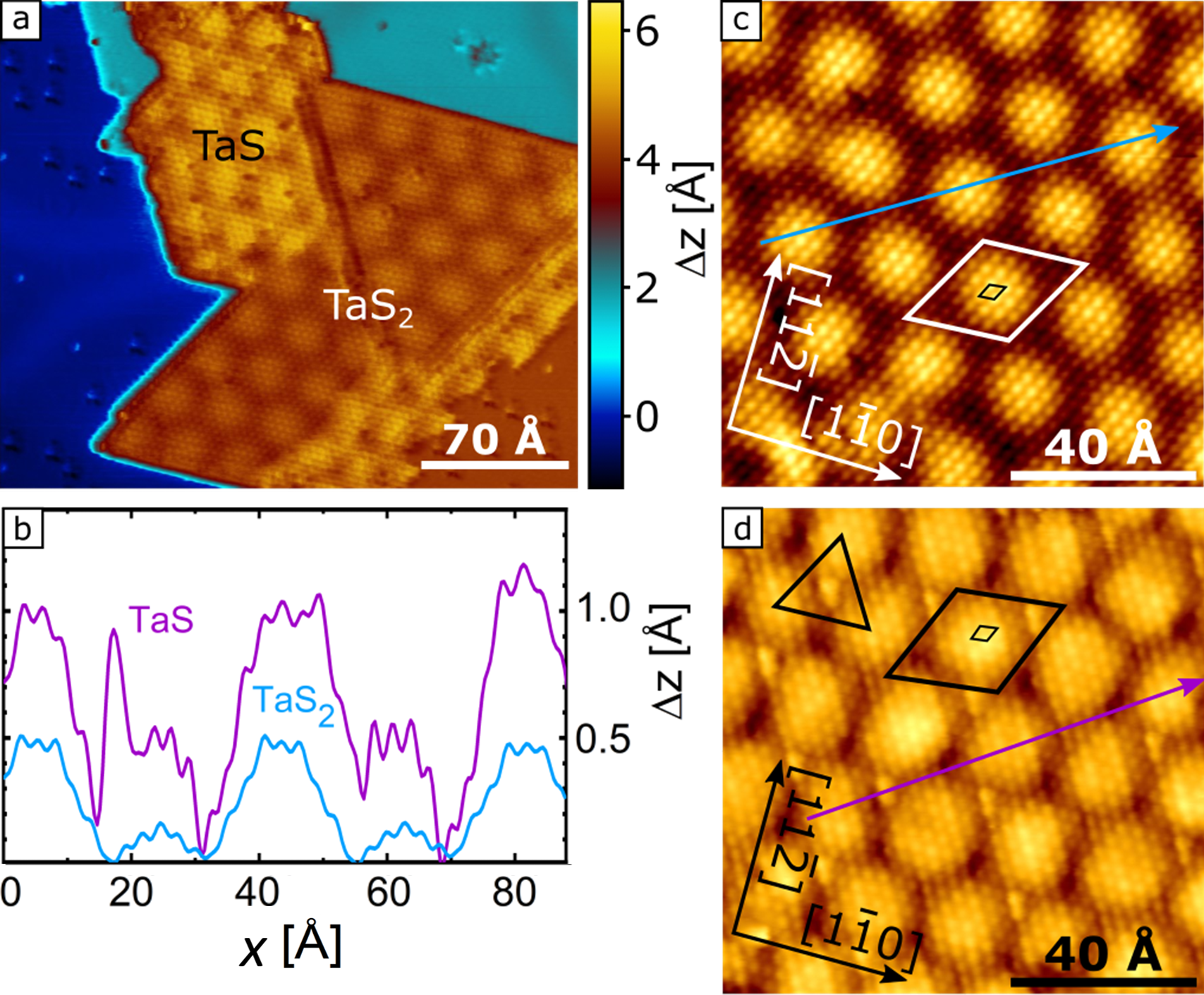}
	\caption{Structure of TaS$_2$ and TaS.
	\textbf{(a)} STM image of a heterostructure island formed by TaS$_2$ (brown) and TaS (yellow) on Au(111) (blue/cyan) ($U=-0.2$~V, $I=3.6$~nA) ($P_{\rm H_2S}=10^{-6}$~mbar, second heating step to 500~K for 20~min without H$_2$S). \textbf{(b)} Line profiles of moir\'{e} corrugation measured in c, d along the lines in the corresponding colours.
	Atomic resolution STM image of \textbf{(c)} TaS$_2$ ($U=-0.5$~V, $I=1.6$~nA) ($P_{\rm H_2S}= 10^{-6}$~mbar, cooling without H$_2$S) and \textbf{(d)} TaS ($U=-0.4$~V, $I=1.9$~nA) ($P_{\rm H_2S}= 10^{-7}$~mbar). The moir\'{e} and atomic unit cells are marked by the large and small diamond, respectively. The orientation of the gold substrate is indicated. Note that for all STM images shown in this work the Au [$11\bar{2}$] direction always points roughly upwards.
}
		\label{fig:Results-TaS2-Morphology-LatticeStructure}
\end{figure}

Figure \ref{fig:Results-TaS2-Morphology-LatticeStructure} (a) shows an STM image of a sample where both TaS$_2$ and TaS are present. We choose the color scale in a way that the gold substrate appears in shades of blue and TaS$_{2}$ and TaS islands appear brown and yellow, respectively. Due to the lattice mismatch with the gold substrate both materials show a hexagonal moir\'{e} pattern. In this image, TaS$_{2}$ and TaS islands are joint, forming a 2D heterostructure.

%
Figures \ref{fig:Results-TaS2-Morphology-LatticeStructure} (c) and (d) show high-resolution STM images of TaS$_{2}$ and TaS, respectively. Two periodicities are visible: the moir\'{e} periodicity and the much smaller periodicity of the atomic lattice. The unit cell of each lattice is indicated by large and small diamonds. Our previous work indicates that the bright protrusions of the atomic lattice are the S atoms \cite{Dombrowski2021}.

The moir\'{e} structure arises from the variation of the registry between the Ta and S atoms on one hand and the Au substrate atoms on the other hand throughout the supercell, as discussed in detail in Ref.~\cite{Silva2021}. This is neglected in the schematic sketches shown in figure~\ref{fig:TaS2Au111stackings}, where we depict the tantalum sulfides in a simplified $(1 \times 1)$ superstructure. In this sketch, we select the local registry where the interaction between  tantalum sulfide and Au(111) is strongest, which is found for the lower S on top of Au for the case of TaS$_2$ and for Ta in a threefold-hollow site for the case of TaS. Note that handily this leads to the same schematic top view for both phases (figure~\ref{fig:TaS2Au111stackings} (a)). This specific registry is found in the regions that give rise to the bright maxima in the moiré cell \cite{Dombrowski2021}. 

TaS$_{2}$ is very well ordered and uniform, exhibiting a smooth moir\'{e} modulation with a pronounced maximum in the center of the unit cell framed by two minima (figure \ref{fig:Results-TaS2-Morphology-LatticeStructure} (c)). The moiré lattice is highly aligned with the TaS$_2$ lattice, which proves that the dense-packed rows of TaS$_2$ and Au(111) are aligned. A moiré analysis yields a lattice constant of $3.30 \pm 0.01$~{\AA} \cite{Dombrowski2021}. The moir\'{e} pattern induces an average apparent corrugation of $\Delta h_{\rm{TaS}_2}=0.4 \pm 0.1$~{\AA} under the tunneling conditions used here  (figure \ref{fig:Results-TaS2-Morphology-LatticeStructure} (b)). The smooth variation and the low amplitude indicate a weak interaction with the substrate.

In contrast, the structure of TaS is more diverse and we can clearly distinguish three moir\'{e} areas (figure \ref{fig:Results-TaS2-Morphology-LatticeStructure} (d)): a pronounced broad primary maximum in the center of the unit cell, a pronounced minimum (left part of the unit cell) and secondary bright area (right part of the unit cell), which often but not always exhibits one or more particularly bright spots. Also here the dense-packed rows of overlayer and substrate are aligned, the lattice constant is $3.32 \pm 0.01$~{\AA} \cite{Dombrowski2021}. For the overall apparent corrugation we find $\Delta h_{\rm{TaS}}=0.9 \pm 0.2$~{\AA} (figure \ref{fig:Results-TaS2-Morphology-LatticeStructure} (b)). The exact values of the moir\'{e} corrugation depend on the tunnel parameters, but TaS is always rougher than TaS$_{2}$. Usually, the moir\'{e} corrugation of TaS is approximately twice as large as the moir\'{e} corrugation of TaS$_{2}$. The more rough structure of TaS stems from a strong interaction with the Au surface.

Using DFT, we calculated the partial charges per tantalum sulfide unit cell of the individual layers (upper S, Ta, lower S, Au(111)) for both TaS$_2$ and TaS, see table~\ref{tab:charge}. For comparison, we include freestanding TaS$_2$. TaS$_2$ adsorbed on Au(111) is remarkably similar to freestanding TaS$_2$. It is also noteworthy that with respect to charge transfer the upper S layer in TaS is almost identical to TaS$_2$, which can explain the similar appearance in STM. Even though the partial charge of the Ta layer is different for the two sulfides, this difference is modest considering the fact that for TaS$_2$ the Ta is covalently bound to six S, whereas for TaS it is only bound to three S and the substrate. The strongest difference is found for the partial charge of Au(111), which even changes sign. 

\begin{table}
\centering
\begin{tabular}{l|r|r|r|r}
	&	\multicolumn{1}{c|}{upper S}	& \multicolumn{1}{c|}{Ta} & \multicolumn{1}{c|}{lower S} & \multicolumn{1}{c}{Au}  \\
\hline
freestanding TaS$_2$	& -0.82	& +1.93	& -0.82	& n.a. \\
TaS$_2$/Au(111)				& -0.83	& +1.60 & -0.84	& +0.07 \\
TaS/Au(111)						& -0.86	& +1.44	& n.a.	& -0.58 \\
\end{tabular}
\caption{Partial charge of the atoms in the individual layers in units of the elementary charge $e$ (n.a. = not applicable.}
\label{tab:charge}
\end{table} 

\label{sec:Orientation}

\begin{figure}
	\centering
		\includegraphics[width=\columnwidth]{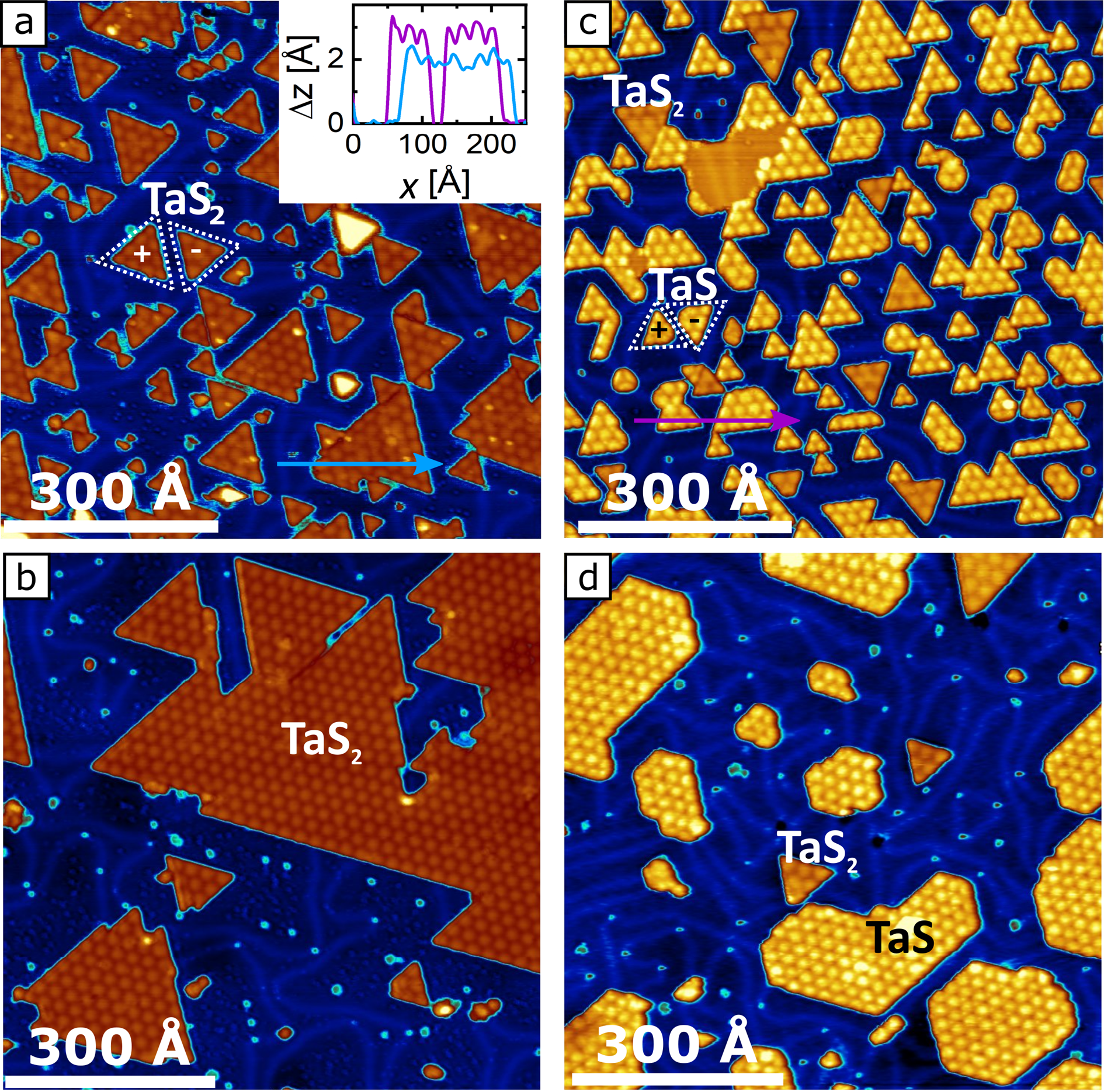}
	\caption{Island shape and orientation of TaS$_2$ and TaS.
	\textbf{(a)} Small islands of TaS$_2$ ($R_{\rm{Ta}} = 0.06\,\rm{ML/min}$, $\Theta_{\rm{Ta}} = 0.24\,\rm{ML}$, $P_{\rm{H}_2\rm{S}} = 10^{-6}$~mbar), ($U=-1.2$~V, $I=110$~pA).	The two dashed triangles marked + and - exemplify the two possible orientations (also in (c)). Inset: Line profiles of TaS$_2$ (blue) and TaS (violet) islands along the lines marked in (a,c) using the same colours. \textbf{(b)} Large islands of TaS$_2$ ($R_{\rm{Ta}} = 0.06\,\rm{ML/min}$, $\Theta_{\rm{Ta}} = 0.24\,\rm{ML}$, $P_{\rm{H}_2\rm{S}} = 10^{-6}$~mbar), ($U= -0.6$~V, $I=450$~pA). 	\textbf{(c)} Small islands of TaS ($R_{\rm{Ta}} = 0.05\,\rm{ML/min}$, $\Theta_{\rm{Ta}} = 0.26\,\rm{ML}$), ($U=-1.6$~V, $I=470$~pA).
	\textbf{(d)} Large islands of TaS ($R_{\rm{Ta}} = 0.06\,\rm{ML/min}$, $\Theta_{\rm{Ta}} = 0.24\,\rm{ML}$), ($U=-0.3$~V, $I=110$~pA).
	}
	\label{fig:Results-TaS2-Morphology-Islandshape}
\end{figure}

TaS$_2$ islands on Au(111) prefer a triangular shape, as shown in figure \ref{fig:Results-TaS2-Morphology-Islandshape} (a, b). The edges of the triangles are parallel to the dense packed atomic rows. We find two orientations that we denote (+) and (-), which differ by a rotation of $180^{\circ}$. For (+), one tip of the triangular shaped island points along the $[11\bar{2}]$-direction of the gold substrate, the other tips point along $[\bar{2}11]$ and $[1\bar{2}1]$. More formally: A vector from the midpoint between two vertices to the third vertex points along $[11\bar{2}]$. For (-), the triangles  point along $[\bar{1}\bar{1}2]$ (and $[2\bar{1}\bar{1}]$, and $[\bar{1}2\bar{1}]$).

Also small TaS islands (figure \ref{fig:Results-TaS2-Morphology-Islandshape} (c)) have a triangular shape, allowing us to use the classification (+)/(-) for this phase as well. Large TaS islands have a more hexagonal shape (figure \ref{fig:Results-TaS2-Morphology-Islandshape} (d)).

For TaS there is another indicator of threefold symmetry: in figure~\ref{fig:Results-TaS2-Morphology-LatticeStructure} (d) we always find the secondary maximum in an up-pointing triangle formed by the primary maxima (marked in the figure). From small triangular TaS islands as those shown in figure~\ref{fig:Results-TaS2-Morphology-Islandshape} (c) we can deduce that this is a feature of the (+)-orientation, while for the (-)-orientation the situation is reversed. This allows us to determine the orientation also for hexagonally or irregularly shaped TaS islands as those in figure~\ref{fig:Results-TaS2-Morphology-Islandshape} (d).

We propose that the two different orientations that we determine phenomenologically from the alignment of triangular islands with respect to the substrate are due to two different orientations of the tantalum sulfide lattice linked by a $180^{\circ}$-rotation as indicated in figure~\ref{fig:TaS2Au111stackings}. This will be analysed in full detail further below.

\label{sec:Stacking}

\begin{figure}
	\centering
		\includegraphics[width=0.5\textwidth]{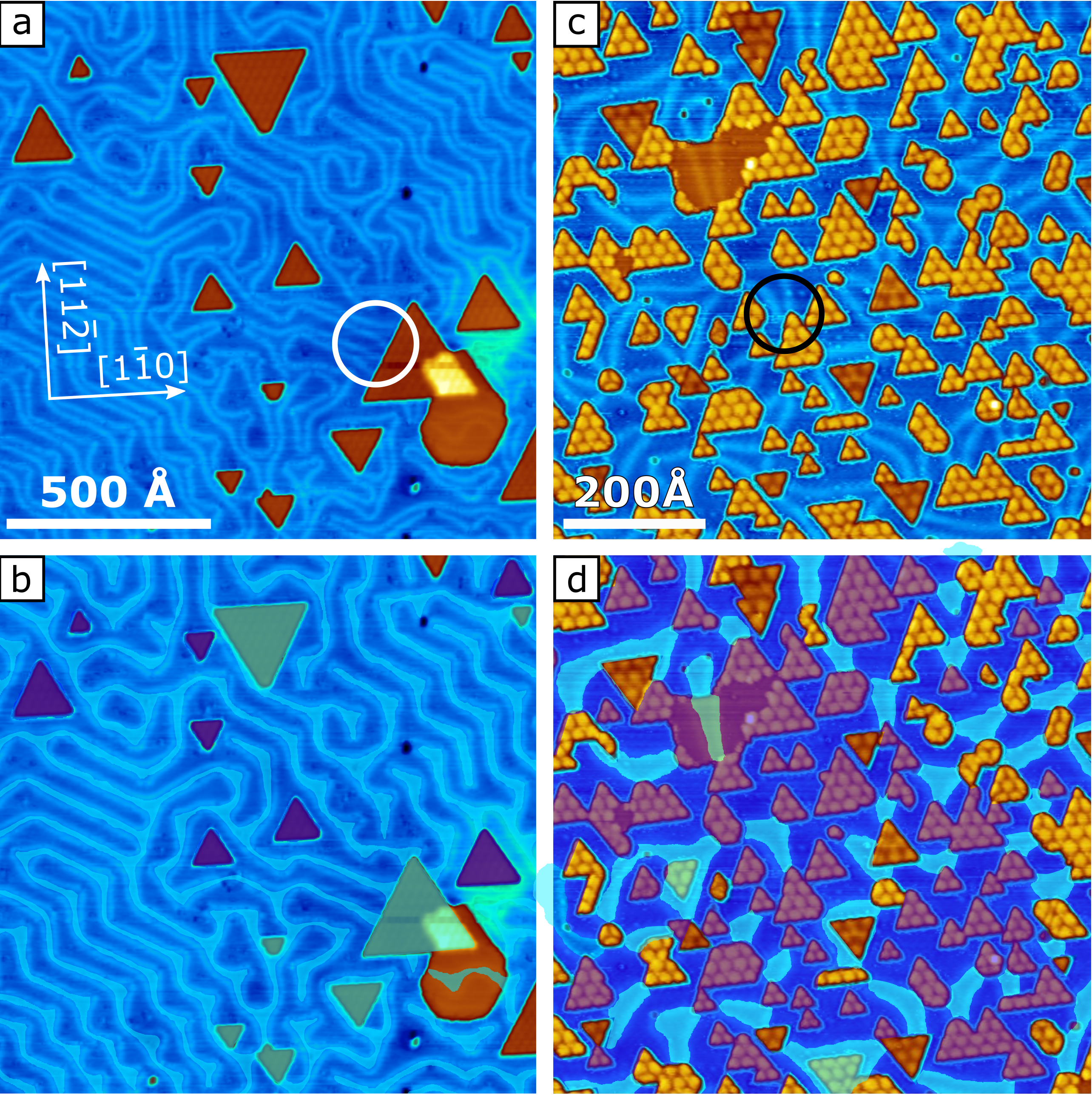}
	\caption{Correlation of tantalum sulfide orientation and Au surface stacking.
	STM image of \textbf{(a)} TaS$_2$-(+)- and (-)-islands ($\Theta_{\rm{Ta}} = 0.1\,\rm{ML}$, $P_{\rm{H}_2\rm{S}} = 10^{-7}$~mbar), ($U=-1$~V, $I=180$~pA),	and \textbf{(c)} (same as \ref{fig:Results-TaS2-Morphology-Islandshape} c) TaS-(+)- and (-)-islands on Au(111) (few TaS$_2$ islands are also present) with the herringbone pattern visible ($R_{\rm Ta} = 0.05$~ML/min, $\Theta_{\rm{Ta}} = 0.26\,\rm{ML}$), ($U=-1.6$~V, $I=470$~pA).	Circles mark special features of the reconstruction lines close to island edges, see text. \textbf{(b,d)} Same STM image as in (a,c), but the tantalum sulfide islands and the herringbone pattern are coloured in blue for fcc and cyan for hcp Au surface stacking. Islands that cannot be unambiguously matched, because they are at the border of the image or do not have a clear triangular shape are not coloured.
	}
	\label{fig:Results-TaS2-Morphology-Herringbone}
\end{figure}

We can sort the tantalum sulfide islands also with respect to the stacking of the Au(111)-substrate. In several of our STM images, the characteristic Au(111)-herringbone reconstruction \cite{Barth1990} is visible as a modulation of the blue background. Locally, this reconstruction has the form of parallel bright lines. The majority of the gold atoms is present in the conventional fcc-stacking (larger space between two bright lines), but for a minority, there is a stacking-fault between the topmost and the second layer which induces hcp-type stacking (smaller space between two bright lines).

By analysing the herringbone pattern near tantalum sulfide islands we can determine the stacking of the Au(111)-substrate underneath the tantalum sulfide islands, analogously to the approach of Krane \emph{et al.} for MoS$_2$/Au(111) \cite{Krane2018}. Figure \ref{fig:Results-TaS2-Morphology-Herringbone} illustrates the matching principle: The original STM-images shown in (a) and (c) are colour-coded in (b) and (d): the fcc areas and correspondingly matched tantalum sulfide islands are marked transparently in dark blue, the hcp areas and matched tantalum sulfide islands in cyan. When the herringbone avoids the island or the lines close at the island edge (usually accompanied by a slight decrease of the hcp herringbone area near the island, see black circle in figure \ref{fig:Results-TaS2-Morphology-Herringbone} (c)), the island grows on fcc-stacked gold. When the herringbone lines surround the island or the hcp region becomes wider near the island edge (see the white circle in figure~\ref{fig:Results-TaS2-Morphology-Herringbone} (a)), the island grows on hcp-stacked gold.

The experimentally observed distribution for the two criteria ((+)- or (-)-orientation, fcc or hcp substrate stacking) is given in Table \ref{tab:StackingTaSx}. We counted several hundred islands, from a full range of preparation conditions. We find an almost equal distribution of the two orientations for TaS$_2$, whereas for TaS the (+)-orientation is strongly favoured. Similarly, we do not observe a strong preference for a specific substrate stacking for TaS$_2$, but TaS islands almost exclusively grow on fcc-stacked Au(111). Taken together, we find that the orientation of both tantalum sulfide phases is linked to the Au(111) surface stacking, the (+)-orientation prefers to grow on fcc-stacked Au areas, whereas the {(-)}-orientation prefers to grow on the hcp-stacked areas. For TaS, this preference is much more pronounced. The fact that the substrate can imprint its orientation on the TaS much better than on TaS$_2$ indicates that TaS interacts more strongly with its support.

\begin{table}
\centering
\begin{tabular}{c|c||r|r|r}
  & & \multicolumn{1}{c|}{fcc}	& \multicolumn{1}{c|}{hcp} & \multicolumn{1}{c}{sum} \\
\hline
\hline
  \multirow{3}{*}{TaS$_2$} 	& (+) &	0.42 				& 0.03 				& {\bf 0.45} \\
														& (-) & 0.11 				& 0.44 				& {\bf 0.55} \\
														& sum & {\bf 0.53}	& {\bf 0.47}	&            \\
\hline
  \multirow{3}{*}{TaS} 			& (+) & 0.95 				& 0.00 				& {\bf 0.95} \\
														& (-) & 0.00				& 0.05 				& {\bf 0.05} \\
														& sum	& {\bf 0.95}	& {\bf 0.05}	& \\

\end{tabular}
\caption{Correlation of tantalum sulfide orientation {((+)/(-))} and Au(111) substrate stacking (fcc/hcp). 311 islands were analysed for TaS$_2$ and 278 for TaS.}
\label{tab:StackingTaSx}
\end{table}

The presence of more than one orientation is detrimental for the application of averaging techniques as angle-resolved photoemission spectroscopy (ARPES) or surface X-ray diffraction (SXRD), as then a superposition of different structures is observed. This can even restore inversion symmetry globally, making special features in the band structure like the opposite spin polarization at the K and K' points invisible. Growth of single-oriented ultrathin layers is therefore highly desirable. For the related systems MoS$_2$/Au(111) and WS$_2$/Au(111), recipes that lead to a unique orientation have been empirically found \cite{Bana2018,Krane2018,Tumino2019,Beyer2019,Bignardi2019}.

Our growth process highly favours the (+)-orientation for TaS, whereas TaS$_2$ is equally present in both orientations. In the following, we will demonstrate that we can exploit the unique orientation of TaS to reach nearly perfect alignment of TaS$_2$.

\begin{figure}
	\centering
		\includegraphics[width=\columnwidth]{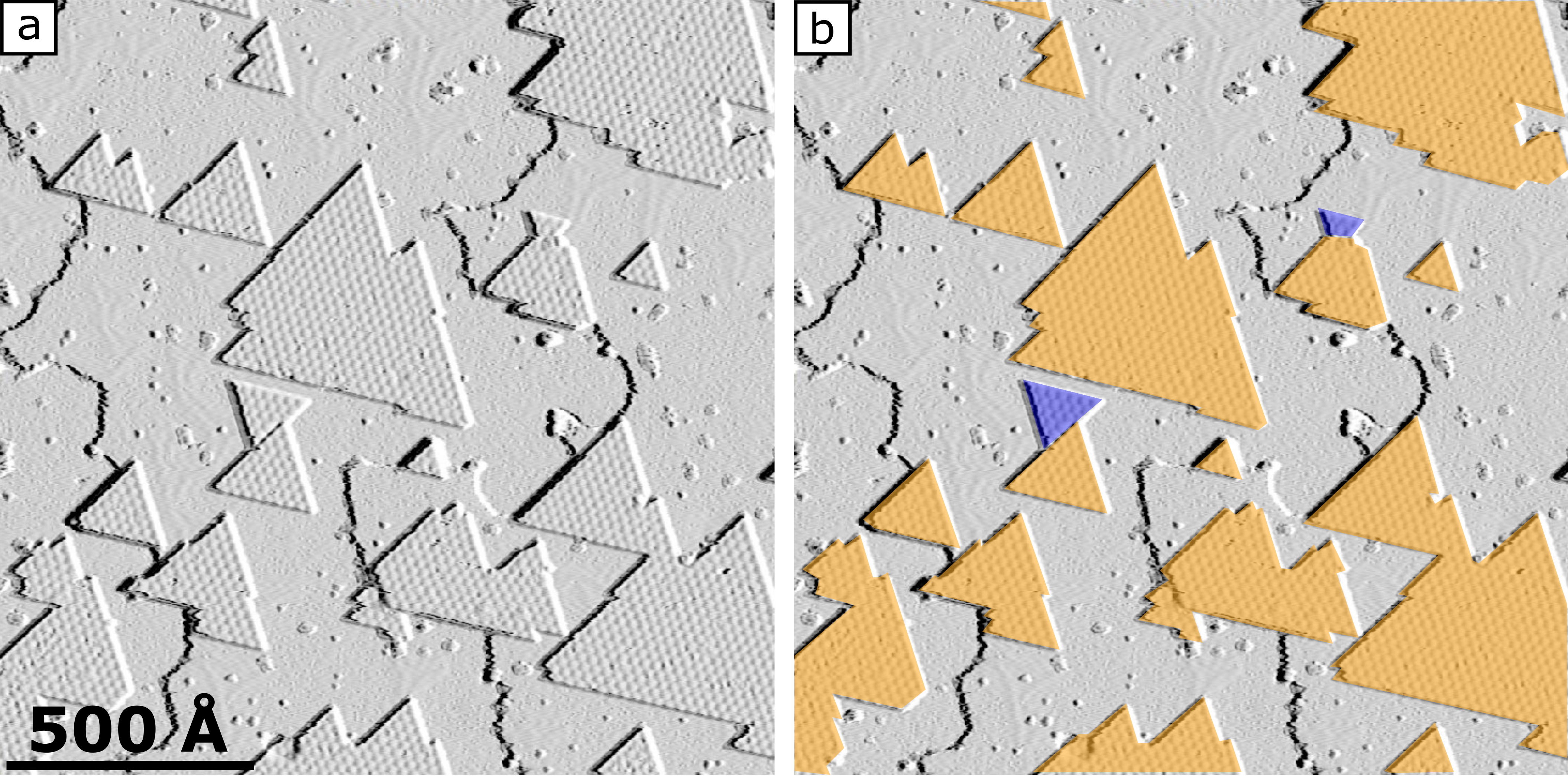}
	\caption{Two-step growth process. 
	\textbf{(a)} STM image of TaS$_2$ grown by the two-step growth process. The STM image has been differentiated to enhance the edge contrast and emphasize the TaS$_2$ islands. The majority of islands is present in the (+)-orientation. 
	\textbf{(b)} Same image as in (a), the orientation of the islands is marked in yellow ((+)) and blue ((-)), ($U=-0.4$~V, $I=550$~pA).
	}
	\label{fig:Results-TaS2-TwoStepGrowth}
\end{figure}
We employ a two-step growth process: (i) Growth of TaS ($R_{\rm Ta} = 0.08$~ML/min, $\Theta_{\rm{Ta}} = 0.25\,\rm{ML}$, $P_{\rm{H}_2\rm{S}} = 10^{-9}$~mbar, $T = 800$~K). (ii) Transformation of TaS into TaS$_2$ by annealing in a high H$_2$S atmosphere ($P_{\rm{H}_2\rm{S}} = 10^{-6}$~mbar, $T = 800$~K). The result of this two step growth process is shown in figure \ref{fig:Results-TaS2-TwoStepGrowth}. We find extended triangular islands of TaS$_2$. Almost all TaS$_2$ ($\approx 99$~\%) is present in the (+)-orientation.

This result shows that we can engineer phase and alignment of the tantalum sulfide on Au(111). The orientation of TaS is preserved during the phase change into TaS$_2$. This also indicates that the atomic structure of both phases, including the registry with the substrate, must be closely related. In summary, the growth protocol developed here can be employed in the future for the preparation of TaS$_2$/Au(111) with a unique orientation.

The fact that TaS can be nicely post-sulfurized with key features of the crystal structure staying intact is also interesting with respect to the growth of Janus monolayers of TMDCs where the upper and lower chalcogen layer are different (e.g. Se-Ta-S). There is a lot of interest in this class of 2DMs due to special properties arising from the breaking of in-plane symmetry \cite{Lu2017}. We speculate that in our system a two-step growth process as detailed above where in the second step S is replaced by Se or Te, could be a viable pathway towards well-defined Janus membranes.

Now we will go beyond a phenomenological description and link orientation and stacking of the tantalum sulfides to the atomic registry of the islands with respect to the Au(111) substrate.

In general, when a threefold-symmetric 2D-crystal is placed on a sixfold-symmetric substrate, two different orientations related by a rotation by 180$^{\circ}$ are to be expected. For fcc(111), only the first layer is sixfold-symmetric, the subsequent layers reduce this to three-fold symmetry. This can lead to an energetic difference between the two orientations which can result in a reduction or even full suppression of one variant, as has been observed previously for WS$_2$/Ag(111) \cite{Ulstrup2017} or hBN/Ir(111) \cite{Orlando2014,FarwickZumHagen2016}.

For Au(111), there is the additional effect of the two different stackings of the substrate. When one only takes the fist two atomic layers into account, the hcp-stacking corresponds to the fcc-stacking with a 180$^{\circ}$-rotation. Therefore, the (+)-orientation on fcc is closely related to the (-)-orientation on hcp, and the (-)-orientation on fcc is closely related to the (+)-orientation on hcp. This is well reflected by our findings summarized in Tab.~\ref{tab:StackingTaSx}: the fact that the sum of entries on the main diagonal is different from the one on the antidiagonal for both TaS$_2$ and TaS is a clear indication for a significant difference between the dissimilar pairs.

In the following, we focus on the Au(111) fcc stacking, as for TaS$_2$ it is not really necessary to distinguish between the stackings, and for TaS the fcc stacking is the only relevant one. We use the simplified scheme in figure~\ref{fig:TaS2Au111stackings}. The two possible registries are labelled according to the relative position of Ta and S with respect to the Au substrate as Ta$_{\rm fcc}$S$_{\rm top}$ and Ta$_{\rm hcp}$S$_{\rm top}$. In explicit, \emph{top} means that the atoms sit directly on top of an Au in the topmost substrate layer, \emph{fcc} indicates that the atoms sit at threefold-hollow sites of fcc-type, and correspondingly \emph{hcp} indicates that the atoms sit at threefold-hollow sites of hcp-type. Conveniently, this nomenclature works for both TaS$_2$ and TaS, one just has to keep in mind that for TaS$_2$, a green circle in figure~\ref{fig:TaS2Au111stackings} represents two S, whereas for TaS it represents just one.

The tantalum sulfide islands have two different kinds of edges, the tantalum-terminated Ta-edge and the sulfur-terminated S-edge. Note that we use these terms purely geometrically (i. e. referring to the ideal truncation of the 2D layer) and not chemically. It is well-known that the edges of 2D sulfides are prone to adsorption and reconstruction \cite{Lauritsen2007}, which is, however, not relevant for our discussion here.  In figure \ref{fig:TaS2Au111stackings} we depict the islands as regular hexagons, where Ta- and S-edges have the same length. However, in experiment, this will only be found in special cases, for example, when the islands are in thermodynamic equilibrium with their surroundings and the edge free energies of Ta- and S-edges are the same. If one edge is kinetically or energetically preferred, the islands become more triangular, as indicated in figure \ref{fig:TaS2Au111stackings} by the blue triangles for the assumption of a preference for the Ta-edge. The fact that we observe triangles for both TaS (for small islands) and TaS$_2$ (always) directly indicates that one edge type is indeed preferred. The orientation ((+) or (-)) of the islands is thus directly related to their registry (Ta$_{\rm fcc}$S$_{\rm top}$ or Ta$_{\rm hcp}$S$_{\rm top}$.) In the following, we want to find this relation, which proves to be notoriously difficult.

Our DFT-calculations show for both phases that Ta$_{\rm fcc}$S$_{\rm top}$ is energetically preferred (by 2.9~meV and 6.4~meV per tantalum sulfide unit cell for TaS$_2$ and TaS, respectively). Under the assumption that our experimentally observed preference of one orientation is (at least partially) due to energetics, this implies that the (+)-orientation is Ta$_{\rm fcc}$S$_{\rm top}$ and the (-)-orientation is Ta$_{\rm hcp}$S$_{\rm top}$. There are surely factors beside energetics that determine the island orientation and could in principle even reverse the relation. However, two arguments speak against this in our case: (i) the relation holds for both phases despite their obviously different growth behaviour and (ii) for TaS the energy difference is larger in line with the stronger preference for one orientation found experimentally.

Martincova \emph{et al.} calculated the energetically preferred shapes of freestanding TaS$_2$ islands under variation of the S chemical potential \cite{Martincova2020}. Under S-poor conditions, they find a triangular shape exposing solely the Ta-edge. We propose that our growth conditions are indeed S-poor, because our TaS$_2$ islands coexist with the S-deficient TaS-phase. In consequence, our triangular TaS$_2$ islands should be Ta-terminated, as tentatively assumed in figure~\ref{fig:TaS2Au111stackings}. This again implies that the (+)-orientation is Ta$_{\rm fcc}$S$_{\rm top}$ and the {(-)}-orientation is Ta$_{\rm hcp}$S$_{\rm top}$ 

Based on these two pieces of evidence (energetically preferred Ta$_{\rm fcc}$S$_{\rm top}$ registry, preferred Ta-edge), in the following, we assume that on fcc-stacked Au(111) the  (+)-orientation is Ta$_{\rm fcc}$S$_{\rm top}$ and the {(-)}-orientation is Ta$_{\rm hcp}$S$_{\rm top}$. For hcp-stacked Au(111) this relation is reversed. Note that for hcp stacking our nomenclature is used a little bit sloppy: One threefold-hollow side has a Au atom directly below, so we can label this hcp, while the other has a Au atom \emph{three} layers below due to the stacking fault, but we still label this as fcc.

In our previous paper~\cite{Dombrowski2021} we chose to display the relaxed structure from DFT for the Ta$_{\rm hcp}$S$_{\rm top}$ registry for both TaS$_2$ and TaS in lack of better knowledge on the preferred registry (which turns out to be the opposite one). However, this does not affect any arguments given in this earlier paper.

A slight inconsistency between experiment and theory remains for the case of TaS: The secondary bright area in the moiré unit cell described above (see figure~\ref{fig:Results-TaS2-Morphology-LatticeStructure} (d)) is found in an up-pointing triangle formed by the primary maxima in experiment. We simulated STM images of this phase (not shown, highly similar to those shown in Ref.~\cite{Dombrowski2021} for the Ta$_{\rm hcp}$S$_{\rm top}$-registry except for a $180^{\circ}$-rotation). These simulations show a feature of similar appearance, but for the Ta$_{\rm fcc}$S$_{\rm top}$-registry this is located in the down-pointing triangles. However, we consider this as a minor problem due to the well-known fallacies of STM simulations (tunneling parameters; tip state) and the potentially simplified geometry we used for the S-poor tantalum sulfide phase in the calculations.

\subsection{Boundaries}
\label{sec:Boundaries}

\begin{figure}
	\centering
		\includegraphics[width=\columnwidth]{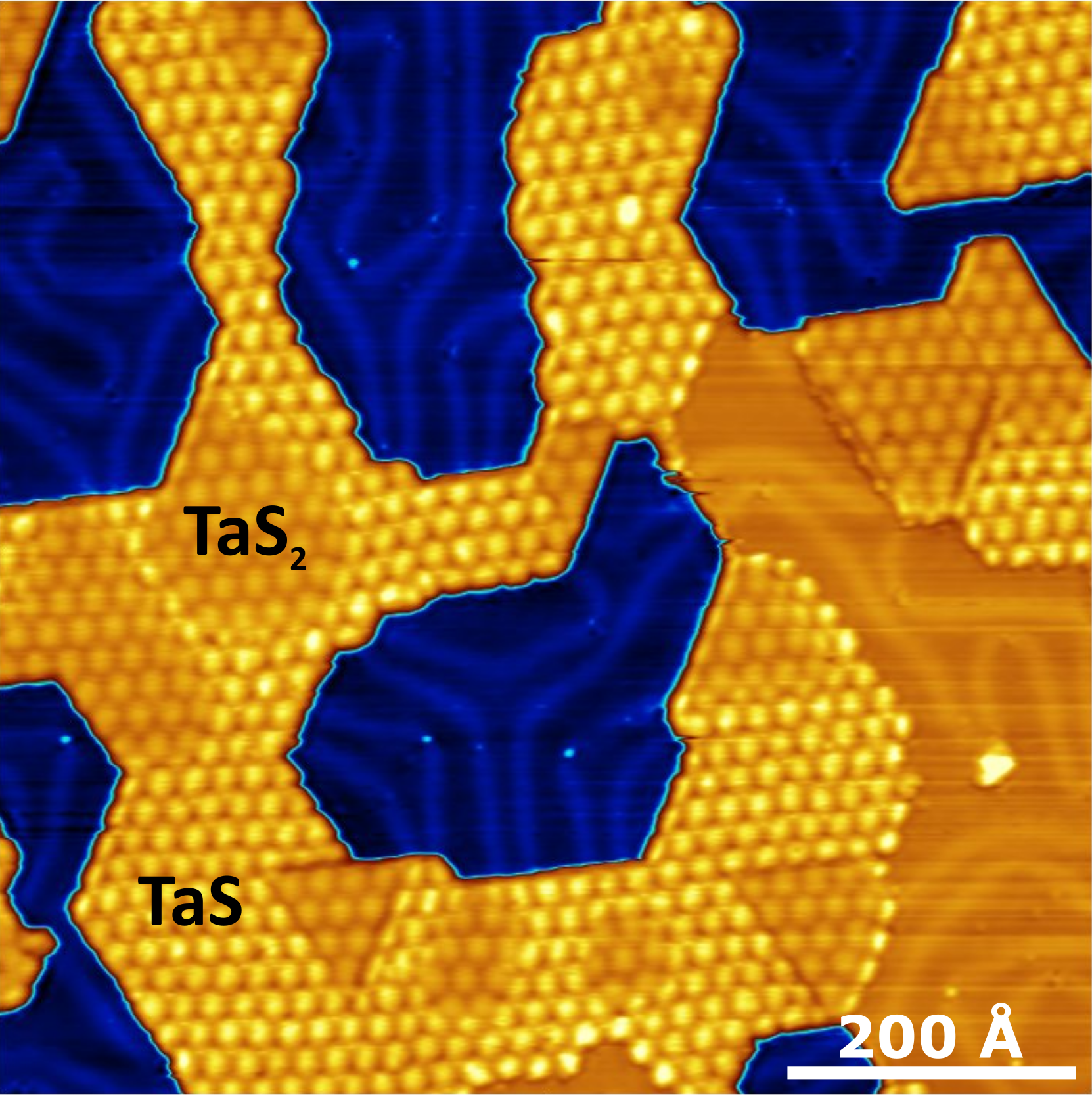}
	\caption{Lateral heterostructure of TaS$_{2}$ and TaS on Au(111) ($P_{\rm{H}_2\rm{S}} = 10^{-7}$~mbar), ($U=-1$~V, $I=330$~pA).}
	\label{fig:Results-TaS2-Heterostructure}
\end{figure}

TaS$_2$ and TaS can form lateral heterostructures as already visible in figure~\ref{fig:Results-TaS2-Morphology-LatticeStructure} (a). Especially for growth with a high amount of Ta at elevated temperatures, these heterostructures are frequently found, see figure \ref{fig:Results-TaS2-Heterostructure}, where TaS encompasses TaS$_2$ islands. We observe two different types of boundaries in our system: boundaries between the same phase in different orientations (mirror twin boundary, MTB) and boundaries between different phases (phase boundary, PB). 

\begin{figure}
	\centering
		\includegraphics[width=\columnwidth]{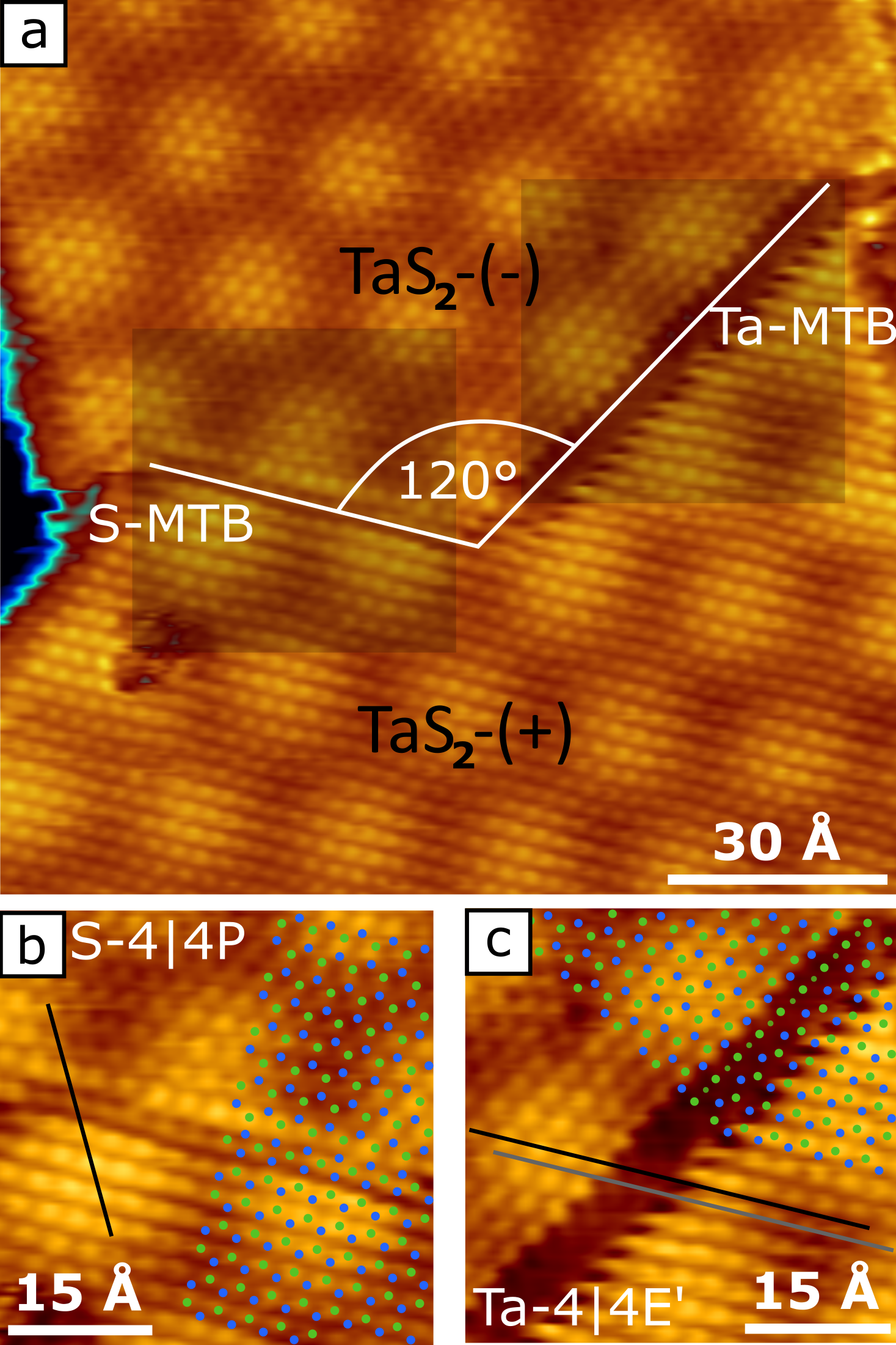}
	\caption{MTB in TaS$_2$.
	\textbf{(a)} STM image of two MTBs in TaS$_2$ ($U=-0.2$~V, $I=1.6$~nA) ($R_{\rm Ta} = 0.08$~ML/min, $\Theta_{\rm Ta}=0.3$~ML, cooling without H$_2$S). \textbf{(b,c)} show a magnification (shaded rectangular areas in (a)) of the S-MTB (superimposed with S-4$|$4P) and the Ta-MTB (superimposed with a model of Ta-4$|$4E'. Circles: S = green (smaller circles indicate S in the bottom layer, almost always eclipsed), Ta = blue. Thin grey / black lines indicate the alignement of atomic rows, see text.}
	\label{fig:Results-TaS2-Boundaries-MTB}
\end{figure}

In figure \ref{fig:Results-TaS2-Boundaries-MTB} (b) an STM image of two joint TaS$_{2}$ islands is shown. The two orientations are separated by two domain boundaries, which follow the dense packed rows of the TaS$_{2}$ atomic lattice. As we do not see a herringbone line emanating from this phase boundary in larger scale images, both islands must reside on the same Au(111) stacking. This is by far the more common situation, and we will focus on this case in the following. Being on the same stacking, the opposite orientation means that the islands have opposite registry. Disregarding the substrate, the two registries are mirror images of each other (see again figure \ref{fig:TaS2Au111stackings}). In consequence, the observed domain boundaries are MTBs \cite{Zande2013,Komsa2017,Batzill2018}. We never observe MTBs for TaS, as it only exists in one orientation on a given Au(111) stacking.

The MTBs in figure~\ref{fig:Results-TaS2-Boundaries-MTB} (a) encompass an angle of $120^{\circ}$, hence one MTB runs along the S- and the other one along the Ta-edge (see again figure~\ref{fig:TaS2Au111stackings}). A large scale image of the same region shows that the upper part has the (-)-orientation, and the lower part has the (+)-orientation, as already suggested by the small part of the edges visible at the left side of figure~\ref{fig:Results-TaS2-Boundaries-MTB} (a). With the assumption of Ta-terminated edges motivated above, the right part of the MTB is therefore between Ta-edges (Ta-MTB), and the left part between S-edges (S-MTB).

 A magnification of the two MTBs with atomic resolution is shown in figures~\ref{fig:Results-TaS2-Boundaries-MTB} (b) and (c). The S-MTB (c) shows atomic contrast within the MTB and the atomic rows are aligned even over the MTB (thin black line). For the Ta-MTB (b) we can not resolve the structure within the MTB and the atomic rows are shifted by approximately half a lattice constant when crossing the MTB (black line vs grey line). 

The most common MTBs are the high symmetry structures 4$|$4P, 4$|$4E and 55$|$8, which usually are observed along the chalcogen edge \cite{Komsa2017,Batzill2018}. Only very few works report an MTB along the metal edge \cite{Wang2020a}. Theoretical calculations of MTBs show that the main structural motifs are very similar for different TMDCs \cite{Lehtinen2015,Zou2015}.

We compare the 4$|$4P, 4$|$4E and 55$|$8 structures to our atomic contrast at the MTBs. For the S-MTB the experimental observations (continued S-rows and atomic resolution in the MTB itself) is reflected best by an S-4$|$4P structure (compare structure sketch in figure~\ref{fig:Results-TaS2-Boundaries-MTB} (b)).  The S-4$|$4P MTB consists of fourfold rings that meet in a point.

For the Ta-MTB (figure~\ref{fig:Results-TaS2-Boundaries-MTB} (c)), we find the best agreement for a theoretical structure calculated by Zou \emph{et al.} \cite{Zou2015} and Lehtinen \emph{et al.} \cite{Lehtinen2015}, which for simplicity we label 4$|$4E' in the following (compare structure sketch in figure~\ref{fig:Results-TaS2-Boundaries-MTB} (c)). This structure induces a shift of S atoms by approximately half a lattice constant, in agreement with our experimental observation. Furthermore, in the  4$|$4E' structure there is a hight difference between the central Ta-layers on both sides, and in consequence also for the upper S-layer. This might explain the lack of atomic resolution in experiment. There is no height difference across the boundary for the S-4$|$4P MTB. The overlay of the structure model correlates with our atomic contrast, assuming that in agreement with STM simulations, the S-atoms appear bright in STM. The position of MTB is thereby chosen in such a way that the last atomic rows imaged correspond to the S-rows that have the normal (binding) environment in the upper S-layer. MTBs along the metal edge are very rare and to the best of our knowledge, none has been reported for pristine, as-grown TMDCs, yet. A 4$|$4E structure along the metal edge has been recently observed for Nb-doped WSe$_2$ \cite{Wang2020a}. The 4$|$4E' MTB is only theoretically found to be stable but has not been observed experimentally.

\begin{figure}
	\centering
		\includegraphics[width=\columnwidth]{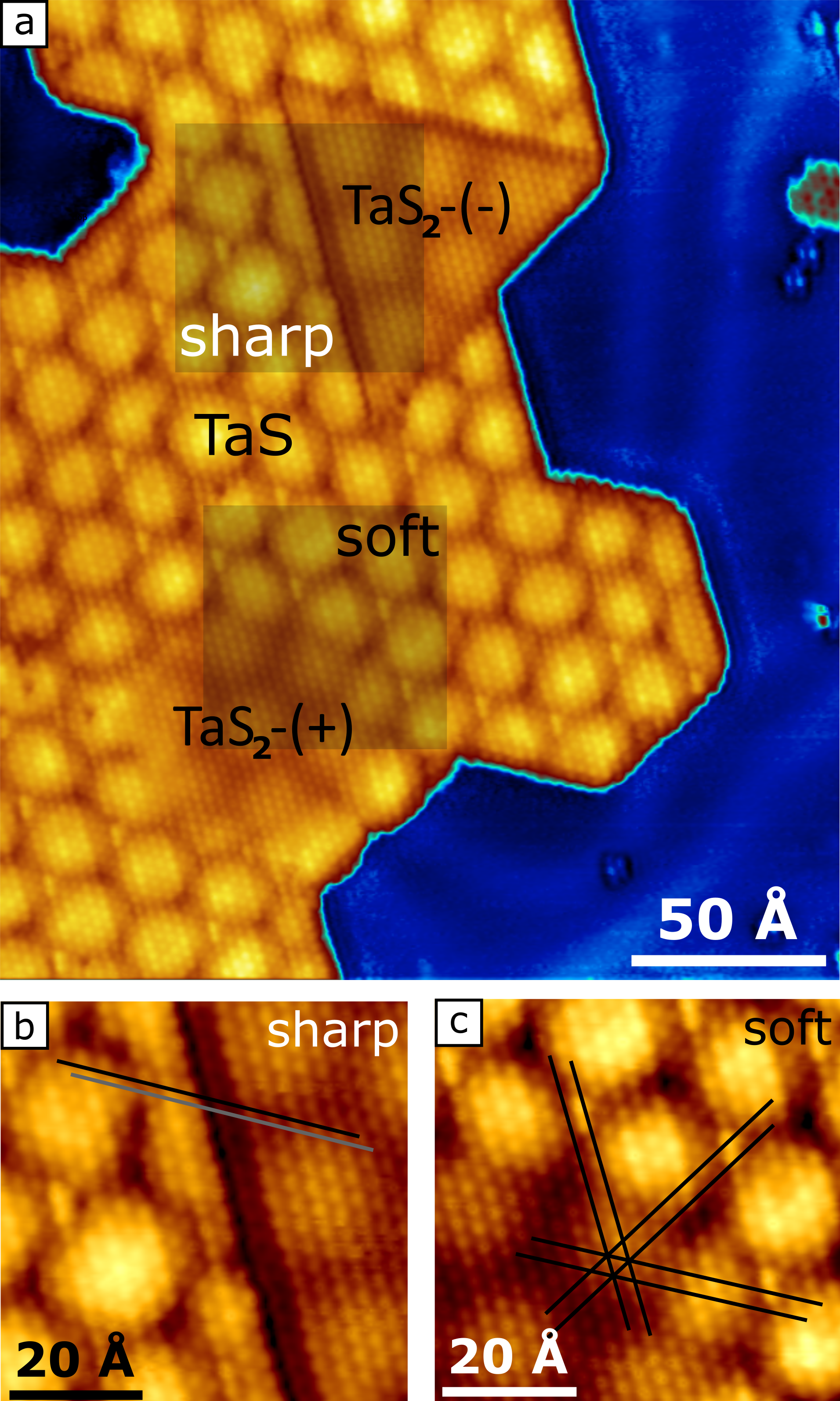}
	\caption{PBs between TaS and TaS$_{2}$. 
	\textbf{(a)} STM image with two TaS$_{2}$ islands embedded in one large TaS island. The two TaS$_{2}$ islands exhibit a different boundary (sharp / soft) with TaS ($U=-0.4$~V, $I=1.9$~nA) ($P_{\rm H_2S} = 10^{-7}$~mbar). 
	\textbf{(b)} and \textbf{(c)} are magnifications of the sharp and soft PB [shaded rectangular areas in (a)], respectively. Thin black /grey lines visualise alignment / misalignment of atomic rows across the boundaries.}
	\label{fig:Results-TaS2-Boundaries-PB}
\end{figure}
In addition to the mirror twin boundaries between TaS$_2$-(+)- and (-)-domains, we observe PBs between TaS and TaS$_{2}$, see the STM image in figure~\ref{fig:Results-TaS2-Boundaries-PB} (a). These PBs come in two varieties that we label \emph{sharp} and \emph{soft}. Soft PBs are always found between TaS$_{2}$ and TaS islands with the same orientation (almost always TaS$_2$-(+) and TaS-(+)). Sharp PBs form exclusively between TaS$_2$ and TaS with different orientation (almost always TaS$_2$-(-) and TaS-(+)). Also here, we have no indication that islands on the different sides of the boundary reside on different Au(111) stacking. In consequence, the soft boundaries are between TaS$_2$ and TaS in the same registry, while the sharp boundaries are between the different phases in opposite registry.

Figure~\ref{fig:Results-TaS2-Boundaries-PB} (a) shows two TaS$_{2}$ islands embedded in one large TaS island. The TaS$_{2}$-(+)-island exhibits a soft boundary along all its edges in touch with TaS, the TaS$_{2}$-(-)-island exhibits a sharp boundary.

The sharp PB can be clearly located (figure \ref{fig:Results-TaS2-Boundaries-PB} (b)). The atomic rows are shifted by approximately half a lattice constant across the boundary (indicated by the thin grey and black lines), similar to what we observed for the Ta-MTB. We interpret the sharp PB as a pseudo-MTB, as here two phases meet that are also mutually rotated by 180$^{\circ}$. It differs from an MTB in the strict sense because of the missing bottom S-layer in TaS.

Figure \ref{fig:Results-TaS2-Boundaries-PB}c shows a magnification of the soft PB with atomic resolution. The boundary is so smooth that it is not even clear where the transition from TaS to TaS$_{2}$ happens exactly. The atomic lattice appears to be completely undisturbed, with continuous alignment of the atomic rows (indicated by the three pairs of parallel black lines). This is true in all directions, hence PBs which form along Ta- and S-edge appear identical in STM. Because of the similar in-plane lattice structure and almost exact match of the lattice parameters of TaS$_{2}$ and TaS the soft PB appears very smooth and exhibits no obvious defects. We can thus engineer heterostructures of different phases with different properties (e. g. one strongly and one weakly anchored to the substrate)  with atomically precise sharp boundaries.

In STM images like figure \ref{fig:Results-TaS2-Heterostructure}  and \ref{fig:Results-TaS2-Boundaries-PB} (a) TaS$_2$ has a lower apparent height than TaS. This finding holds for a wide range of tunneling conditions. It is counter-intuitive in view of our structural model, see figure~\ref{fig:TaS2Au111stackings}. We see two solutions for this puzzle: On one hand, as STM always images a convolution of the geometric and the electronic structure, a possible explanation is the different band structure and resulting different local density of states of TaS$_2$ and TaS \cite{Dombrowski2021}. Also the opposite direction of charging (electrons flow from the substrate to the overlayer for TaS$_2$, but the other direction for TaS, see table~\ref{tab:charge}) can play a role. In addition, TaS is strongly anchored to the substrate, which may lead to a more 3D-character of its electronic states, and thus to a slower decay of the states into the vacuum. On the other hand, it is also possible that our structural model for TaS is incomplete: Maybe the lower S-layer is not completely absent, or partially replaced by Au. The high mobility of Au atoms at the annealing temperatures combined with the strong interaction between TaS and the substrate may even lead to an accumulation of a full extra gold layer below TaS, which would explain why it is imaged higher in STM.

\subsection{Carpet-like growth of TaS$_2$ over Au step edges}

\begin{figure}
	\centering
		\includegraphics[width=\columnwidth]{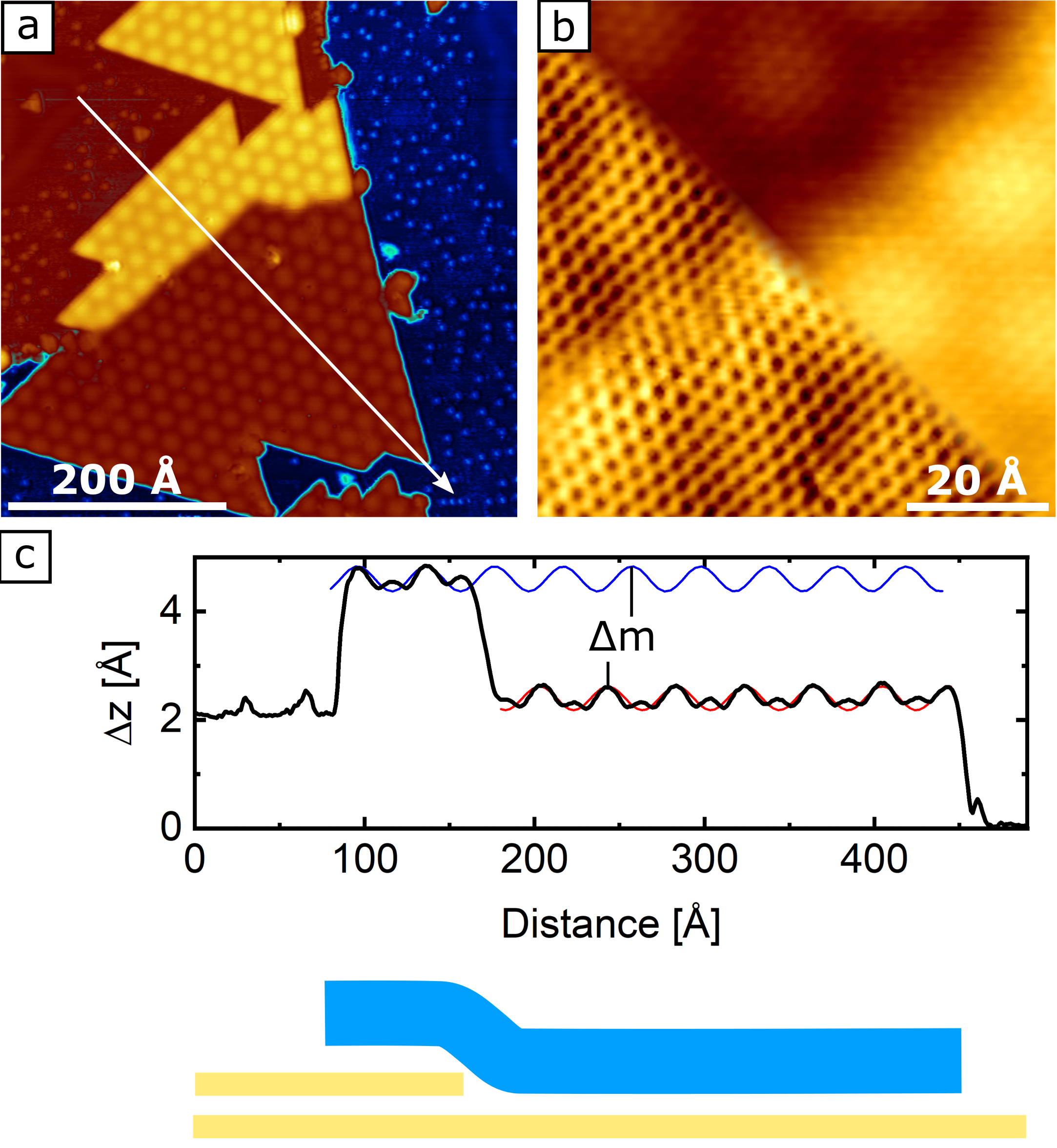}
	\caption{Step edge growth of TaS$_{2}$. 
	\textbf{(a)} STM image of a TaS$_2$ island that covers an Au step edge ($U=-0.7$~V, $I=320$~pA) (second heating step $P_{\rm H_2S} = 10^{-6}$~mbar for 20 min). 
	\textbf{(b)} Atomic resolution of TaS$_2$ covering the Au step, the upper corner of the image shows the topography image of the step edge with both, the moir\'{e} and atomic periodicity resolved. The lower corner shows the same area, but as a differentiated STM image, so that the continuity of the atomic lattice is better visible ($U=-0.5$~V, $I=1.6$~nA) (cooling without H$_2$S). 
	\textbf{(c)} Line profile showing the moir\'{e} shift across the step edge and schematic side view of the structure. Colors: TaS$_2$=blue, Au=yellow.
	}
	\label{fig:Results-TaS2-StepdEdgeGrowth}
\end{figure}
Figure \ref{fig:Results-TaS2-StepdEdgeGrowth} (a) shows a TaS$_{2}$ island that covers a gold step. The bright yellow part of the island is on the upper gold terrace whereas the darker red-brown part is located on the lower gold terrace. We find that this leads to an atomic lattice that is bent but uninterrupted, as visible in the atomically resolved image shown in figure \ref{fig:Results-TaS2-StepdEdgeGrowth} (b). The upper corner of the image shows the topography image of the step edge with both, the moir\'{e} and atomic periodicity resolved. The lower corner shows the same area, but as differentiated STM image, so that the continuity especially of the atomic lattice is better visible.
Such growth is already known for several 2D materials \cite{Coraux2008,Sorensen2014,Dendzik2015}. 

We can determine the bending radius of the 2D-layer when it goes down a substrate step by making use of the magnification effect of the moiré \cite{Coraux2008}. The smaller the bending radius, the larger the relative shift $\Delta_{{\rm TaS}_2}$ of the atoms in the lower part from the positions expected by extrapolation from the upper part. While this alone is very difficult to measure, this shift is reflected in a much larger shift $\Delta_{\rm moir\acute{e}}$  of the moire superstructure. Explicitly, they are linked by the expression $\Delta_{\rm{TaS}_2} = a_{\rm{TaS}_2} ( \Delta_{\rm{Au}}/a_{\rm{Au}} - \Delta_{\rm{moir\acute{e}}}/a_{\rm{moir\acute{e}}} )$ where $\Delta_{Au}$ is the shift of Au substrate atoms across the step and $a_{\rm Au}=2.88$~{\AA} is the Au in-plane lattice constant \cite{Maeland1964}.

We find a moir\'{e} shift of $\Delta_{\rm{moir\acute{e}}} = 15$~{\AA} for \{100\}-microfacetted steps and $\Delta_{\rm{moir\acute{e}}} = -14$~{\AA} for \{111\}-microfacetted steps. This results in a shift of the atoms $\Delta_{{\rm TaS}_2} = - 4$~{\AA} for \{100\} and $\Delta_{{\rm TaS}_2} = + 4$~{AA} for \{111\}. Using the set of two non-linear equations given in Ref.~\cite{Coraux2008}, we can estimate the bending radius of the TaS$_{2}$ sheet to be of the order of 15 - 30~{\AA}. This correlates roughly to the radius of very small group 4 TMDC nanotubes \cite{Nath2001}. 

Because of the asymptotic behaviour of equations towards large bending radii, the error of $R$ is significantly larger towards large values than small values of $R$. Furthermore, the model described in Ref.~\cite{Coraux2008} assumes a symmetric bending over the step edge, which for TaS$_2$ (with S-Ta-S sandwich structure) can only be true for the Ta-layer in the middle, the S-layers cannot bend symmetrically. Our analysis thus provides just a rough estimate of the order of magnitude of the bending radius.

For TaS, we regularly find islands that are attached to a gold step edge (e.g. figure~\ref{fig:Results-TaS2-Heterostructure}) but we do not observe TaS islands that grow over a gold step edge like TaS$_2$ does. This observation is in line with a strong substrate interaction.


\section{Conclusion}

Using the same ingredients, two different phases of monolayer tantalum sulfide can be grown on Au(111). The conventional 2H-TaS$_2$ and the sulfur deficient TaS$_{2-x}$ with $0 \leq x \leq 1$ (called TaS in this paper). For both phases we typically find triangular islands that can be present in two orientations on Au(111) linked by a $180^{\circ}$-rotation and that can be found on both fcc-stacked and hcp-stacked areas within the Au(111) herringbone reconstruction. While 2H-TaS$_2$ does not show a clear preference both for orientation and substrate stacking, TaS prefers to grow on fcc-stacked gold in just one orientation. We can exploit this for phase engineering, where we transform highly oriented TaS by post-sulfurization into highly oriented 2H-TaS$_2$. Thereby, the breaking of inversion symmetry is globally preserved and enables, among other things, the detection of spin-split valleys at the inequivalent K points by ARPES.

Under the assumption that the edges of the triangular islands are terminated by Ta for both 2H-TaS$_2$ and TaS and that the energetically preferred orientations prevail, we can fully determine the registry of both phases with the substrate for all orientations and both stackings. The Ta$_{\rm fcc}$S$_{\rm top}$-registry is preferred for both phases.

For higher coverages we find mirror twin boundaries between differently oriented domains of 2H-TaS$_2$, both along the Ta-edges and the S-edges. We identify them as 4$|$4P for the S-MTB and 4$|$4E' for the Ta-MTB. In addition we find heterostructures between the different phases. When both phases have the same orientation, these boundaries are atomically sharp and smooth. We are thus able to coherently join materials with different properties.

2H-TaS$_2$ shows a carpet-like growth over step edges of Au(111). We determine the bending radius of the monolayer across the step to be of the order of 15 to 30~{\AA}, which relates to thin TMDC nanotubes.

\ack
The research reported in this publication was supported by funding from King Abdullah University of Science and Technology (KAUST). T. Chagas acknowledges financial support from the Alexander von Humboldt foundation.

\section*{References}

\bibliographystyle{unsrt}
\bibliography{Literatur-Heterostructure.bib}

\begin{thebibliography}{10}

\bibitem{Chen2020}
Y.~Chen, Z.~Lai, X.~Zhang, Z.~Fan, Q.~He, C.~Tan, and H.~Zhang.
\newblock Phase engineering of nanomaterials.
\newblock {\em Nat. Rev. Chem.}, 4:243, 2020.

\bibitem{Wang2018}
R.~Wang, Y.~Yu, S.~Zhou, H.~Li, H.~Wong, Z.~Luo, L.~Gan, and T.~Zhai.
\newblock Strategies on {Phase} {Control} in {Transition} {Metal}
  {Dichalcogenides}.
\newblock {\em Adv. Funct. Mater.}, 28:1802473, 2018.

\bibitem{Bonilla2020}
M.~Bonilla, S.~Kolekar, J.~Li, Y.~Xin, P.~M. Coelho, K.~Lasek, K.~Zberecki,
  D.~Lizzit, E.~Tosi, P.~Lacovig, S.~Lizzit, and M.~Batzill.
\newblock {Compositional Phase Change of Early Transistion Metal Diselenide
  (VSe$_2$ and TiSe$_2$) Ultrathin Films by Postgrowth Annealing}.
\newblock {\em Adv. Mater. Interfaces}, 7:2000497, 2020.

\bibitem{Zhao2020}
X.~Zhao, P.~Song, Wang. C., A.~C. Riis-Jensen, W.~Fu, Y.~Deng, D.~Wan, L.~Kang,
  S.~Ning, J.~Dan, T.~Venkatesan, Z.~Liu, W.~Zhou, K.~S. Thygesen, X.~Luo,
  S.~J. Pennycook, and K.~P. Loh.
\newblock {Engineering Covalently Bonded 2D Layered Materials by
  Self-Intercalation}.
\newblock {\em Nature}, 581:171, 2020.

\bibitem{Arnold2018}
F.~Arnold, R.-M. Stan, S.~K. Mahatha, H.~E. Lund, D.~Curico, M.~Dendzik,
  H.~Bana, E.~Travaglia, L.~Bignardi, P.~Lacovig, D.~Lizzit, Z.~Li, M.~Bianchi,
  J.~Miwa, M.~Bremholm, S.~Lizzit, P.~Hofmann, and C.~E. Sanders.
\newblock {Novel Single-Layer Vanadium Sulphide Phases}.
\newblock {\em 2D Mater.}, 5:045009, 2018.

\bibitem{Cheng2018}
F.~Cheng, Z.~Ding, H.~Xu, S.~J.~R. Tan, I.~Abdelwahab, J.~Su, P.~Zhou,
  J.~Martin, and K.~P. Loh.
\newblock {Epitaxial Growth of Single-Layer Niobium Selenides with Controlled
  Stoichiometric Phases}.
\newblock {\em Adv. Mater. Interfaces}, 5:1800429, 2018.

\bibitem{Chhowalla2013}
M.~Chhowalla, H.~S. Shin, G.~Eda, L.~J. Li, K.~P. Loh, and H.~Zhang.
\newblock The chemistry of two-dimensional layered transition metal
  dichalcogenide nanosheets.
\newblock {\em Nat. Chem.}, 5:263--275, 2013.

\bibitem{Sanders2016}
C.~E. Sanders, M.~Dendzik, A.~S. Ngankeu, A.~Eich, A.~Bruix, M.~Bianchi, J.~A.
  Miwa, B.~Hammer, A.~A. Khajetoorians, and P.~Hofmann.
\newblock {Crystalline and Electronic Structure of Single-Layer TaS$_2$}.
\newblock {\em Phys. Rev. B}, 94:081404, 2016.

\bibitem{Xiao2012}
D.~Xiao, G.-B. Liu, W.~Feng, X.~Xu, and W.~Yao.
\newblock {Coupled Spin and Valley Physics in Monolayers of MoS$_2$ and Other
  Group-VI Dichalocogenides}.
\newblock {\em Phys. Rev. Lett.}, 108:196802, 2012.

\bibitem{Jellinek1962}
F.~Jellinek.
\newblock {The System Tantalum-Sulfur}.
\newblock {\em J. Less Common Met.}, 4:9--15, 1962.

\bibitem{Dombrowski2021}
D.~Dombrowski, A.~Samad, C.~Murray, M.~Petrovi\'{c}, P.~Ewen, T.~Michely,
  M.~Kralj, U.~Schwingenschl\"{o}gl, and C.~Busse.
\newblock {Two Phases of Monolayer Tantalum Sulfide on Au(111)}.
\newblock {\em ACS Nano}, 15:13516, 2021.

\bibitem{Silva2021}
C.~C. Silva, D.~Dombrowski, A.~Samad, J.~Cai, W.~Jolie, J.~Hall, P.~T.~P. Ryan,
  P.~K. Thakur, D.~A. Duncan, T.-L. Lee, U.~Schwingenschlögl, and C.~Busse.
\newblock Structure of monolayer {2H}-{TaS}$_2$ on {Au}(111).
\newblock {\em Phys. Rev. B}, 104:205414, 2021.

\bibitem{PhysRevB.59.1758}
G.~Kresse and D.~Joubert.
\newblock From ultrasoft pseudopotentials to the projector augmented-wave
  method.
\newblock {\em Phys. Rev. B}, 59:1758, 1999.

\bibitem{perdew1996generalized}
J.~P. Perdew, K.~Burke, and M.~Ernzerhof.
\newblock Generalized gradient approximation made simple.
\newblock {\em Phys. Rev. Lett.}, 77:3865, 1996.

\bibitem{grimme2006semiempirical}
S.~Grimme.
\newblock {Semiempirical GGA-Type Density Functional Constructed with a
  Long-Range Dispersion Correction}.
\newblock {\em J. Comput. Chem.}, 27:1787, 2006.

\bibitem{tersoff1985theory}
J.~Tersoff and D.~R. Hamann.
\newblock Theory of the scanning tunneling microscope.
\newblock {\em Phys. Rev. B}, 31:805--813, 1985.

\bibitem{Yu2011}
M.~Yu and D.~R. Trinkle.
\newblock Accurate and efficient algorithm for {Bader} charge integration.
\newblock {\em J. Chem. Phys.}, 134:064111, 2011.

\bibitem{Barth1990}
J.~V. Barth, H.~Brune, and G.~Ertl.
\newblock {Scanning Tunneling Microscopy Observations on the Reconstructed
  Au(111) Surface: Atomic Structure, Long-Range Superstructure, Rotational
  Domains, and Surface Defects}.
\newblock {\em Phys. Rev. B}, 42:9307, 1990.

\bibitem{Krane2018}
N.~Krane, C.~Lotze, and K.~J. Franke.
\newblock {Moir\'{e} structure of MoS$_2$ on Au(111): Local structural and
  electronic properties}.
\newblock {\em Surf. Sci.}, 678:136, 2018.

\bibitem{Bana2018}
H.~Bana, E.~Travaglia, L.~Bignardi, P.~Lacovig, C.~E. Sanders, M.~Dendzik,
  M.~Michiardi, M.~Bianchi, D.~Lizzit, F.~Presel, D.~De~Angelis, N.~Apostol,
  P.~K. Das, J.~Fujii, I.~Vobornik, R.~Larciprete, A.~Baraldi, P.~Hofmann, and
  S.~Lizzit.
\newblock Epitaxial growth of single-orientation high-quality {MoS}$_2$
  monolayers.
\newblock {\em 2D Mater.}, 5:035012, 2018.

\bibitem{Tumino2019}
F.~Tumino, C.~S. Casari, A.~L. Bassi, M.~Passoni, and V.~Russo.
\newblock {Pulsed laser deposition of single-layer MoS$_2$ on Au(111): from
  nanosized crystals to large-area films}.
\newblock {\em Nanoscale Adv.}, 1:643, 2019.

\bibitem{Beyer2019}
H.~Beyer, G.~Rohde, A.~Grubi\v{s}ić~\v{C}abo, A.~Stange, T.~Jacobsen,
  L.~Bignardi, D.~Lizzit, P.~Lacovig, C.~E. Sanders, S.~Lizzit, K.~Rossnagel,
  P.~Hofmann, and M.~Bauer.
\newblock 80\% {Valley} {Polarization} of {Free} {Carriers} in {Singly}
  {Oriented} {Single}-{Layer} {WS}$_2$ on {Au}(111).
\newblock {\em Phys. Rev. Lett.}, 123:236802, 2019.

\bibitem{Bignardi2019}
L.~Bignardi, D.~Lizzit, H.~Bana, E.~Travaglia, P.~Lacovig, C.~E. Sanders,
  M.~Dendzik, M.~Michiardi, M.~Bianchi, M.~Ewert, L.~Buß, J.~Falta, J.~I.
  Flege, A.~Baraldi, R.~Larciprete, P.~Hofmann, and S.~Lizzit.
\newblock Growth and structure of singly oriented single-layer tungsten
  disulfide on {Au}(111).
\newblock {\em Phys. Rev. Mater.}, 3:014003, 2019.

\bibitem{Lu2017}
A.-Y. Lu, H.~Zhu, J.~Xiao, C.-P. Chuu, Y.~Han, M.-H. Chiu, C.-C. Cheng, C.-W.
  Yang, K.-H. Wei, Y.~Yang, Y.~Wang, D.~Sokaras, D.~Nordlund, P.~Yang, D.~A.
  Muller, M.-Y. Chou, X.~Zhang, and L.-J. Li.
\newblock Janus monolayers of transition metal dichalcogenides.
\newblock {\em Nature Nanotechnology}, 12:744, 2017.

\bibitem{Ulstrup2017}
S.~Ulstrup, A.~Grubi\v{s}ić~\v{C}abo, D.~Biswas, J.~M. Riley, M.~Dendzik,
  C.~E. Sanders, M.~Bianchi, C.~Cacho, D.~Matselyukh, R.~T. Chapman,
  E.~Springate, P.~D.~C. King, J.~A. Miwa, and P.~Hofmann.
\newblock Spin and valley control of free carriers in single-layer {WS}$_2$.
\newblock {\em Phys. Rev. B}, 95:041405, 2017.

\bibitem{Orlando2014}
F.~Orlando, P.~Lacovig, N.~G. Omiciuolo, L.~Apostol, R.~Larciprete, A.~Baraldi,
  and S.~Lizzit.
\newblock {Epitaxial Growth of a Single-Domain Hexagonal Boron Nitride
  Monolayer}.
\newblock {\em ACS Nano}, 8:12063, 2014.

\bibitem{FarwickZumHagen2016}
F.~H. {Farwick zum Hagen}, D.~M. Zimmermann, C.~C. Silva, C.~Schlueter,
  N.~Atodiresei, W.~Jolie, A.~J. Mart\'{i}nez-Galera, D.~Dombrowski, U.~A.
  Schr\"oder, M.~Will, P.~Lazi\'{c}, V.~Caciuc, S.~Bl\"ugel, T.-L. Lee,
  T.~Michely, and C.~Busse.
\newblock {Structure and Growth of Hexagonal Boron Nitride on Ir(111)}.
\newblock {\em ACS Nano}, 10:11012, 2016.

\bibitem{Lauritsen2007}
J.~V. Lauritsen, J.~Kibsgaard, S.~Helveg, H.~Tops{\o}e, B.~S. Clausen,
  E.~L{\ae}gsgaard, and F.~Besenbacher.
\newblock {Size-dependent structure of MoS$_2$ nanocrystals}.
\newblock {\em Nat. Nanotechnol.}, 2:53, 2007.

\bibitem{Martincova2020}
J.~Martincov{\'a}, M.~Otyepka, and P.~Lazar.
\newblock {Atomic-Scale Edge Morphology, Stability and Oxidation of
  Single-Layer 2H-TaS$_2$}.
\newblock {\em ChemPlusChem}, 85:2557, 2020.

\bibitem{Zande2013}
A.~M. van~der Zande, P.~Y. Huang, D.~A. Chenet, T.~C. Berkelbach, Y.~You, G.-H.
  Lee, T.~F. Heinz, D.~R. Reichman, D.~A. Muller, and J.~C. Hone.
\newblock {Grains and grain boundaries in highly crystalline monolayer
  molybdenum disulphide}.
\newblock {\em Nat. Mater.}, 12:554, 2013.

\bibitem{Komsa2017}
H.-P. Komsa and A.~V. Krasheninnikov.
\newblock {Engineering the Electronic Properties of Two-Dimensional Transition
  Metal Dichalcogenide by Introducing Mirror Twin Boundaries}.
\newblock {\em Adv. Electron. Mater.}, 3:1600468, 2017.

\bibitem{Batzill2018}
M.~Batzill.
\newblock {Mirror twin grain boundaries in molybdenum dichalcogenides}.
\newblock {\em J. Phys. Condens. Matter}, 30:493001, 2018.

\bibitem{Wang2020a}
B.~Wang, Y.~Xia, J.~Zhang, H.-P. Komsa, M.~Xie, Y.~Peng, and C.~Jin.
\newblock {Niobium doping induced mirror twin boundaries in MBE grwon WSe$_2$
  monolayers}.
\newblock {\em Nano Res.}, 13:1889, 2020.

\bibitem{Lehtinen2015}
O.~Lehtinen, H.-P. Komsa, A.~Pulkin, M.~B. Whitwick, M.-W. Chen, T.~Lehnert,
  M.~J. Mohn, O.~V. Yazyev, A.~Kis, U.~Kaiser, and A.~V. Krasheninnikov.
\newblock {Atomic Scale Microstructure and Properties of Se-Deficient
  Two-Dimensional MoSe$_2$}.
\newblock {\em ACS Nano}, 9:3274, 2015.

\bibitem{Zou2015}
X.~Zou and B.~I. Yakobson.
\newblock {Metallic High-Angle Grain Boundaries in Monolayer Polycrystalline
  WS$_2$}.
\newblock {\em Small}, 11:4503, 2015.

\bibitem{Coraux2008}
J.~Coraux, A.~T. N'Diaye, C.~Busse, and T.~Michely.
\newblock {Structural Coherency of Graphene on Ir(111)}.
\newblock {\em Nano Lett.}, 8:565, 2008.

\bibitem{Sorensen2014}
S.~G. S{\o}rensen, H.~G. F{\"{u}}chtbauer, A.~K. Tuxen, A.~S. Walton, and J.~V.
  Lauritsen.
\newblock {Structure and Electronic Properties of \emph{in Situ} Synthesized
  Single-Layer MoS$_2$ on a Gold Surface}.
\newblock {\em ACS Nano}, 8:6788, 2014.

\bibitem{Dendzik2015}
M.~Dendzik, M.~Michiardi, C.~E. Sanders, M.~Bianchi, J.~A. Miwa, S.~S.
  Gr{\o}nborg, J.~V. Lauritsen, A.~Bruix, B.~Hammer, and P.~Hofmann.
\newblock Growth and electronic structure of epitaxial single-layer {WS}$_2$ on
  {Au}(111).
\newblock {\em Phys. Rev. B}, 92:245442, 2015.

\bibitem{Maeland1964}
A.~Maeland and T.~B. Flanagan.
\newblock Lattice spacings of gold–palladium alloys.
\newblock {\em Can. J. Phys.}, 42:2364, 1964.

\bibitem{Nath2001}
M.~Nath and C.~N.~R. Rao.
\newblock {New Metal Dislufide Nanotubes}.
\newblock {\em J. Am. Chem. Soc.}, 123:4841, 2001.

\end{thebibliography}

\end{document}